\documentstyle{cupconf}

\ifoldfss
\else
  \ifnfssone
    \newmathalphabet{\mathit}
      \addtoversion{normal}{\mathit}{cmr}{m}{it}
      \addtoversion{bold}{\mathit}{cmr}{bx}{it}
    \newmathalphabet{\mathcal}
      \addtoversion{normal}{\mathcal}{cmsy}{m}{n}
    \else
    \ifnfsstwo
    \fi
  \fi
\fi

\def\hexnumber#1{\ifcase#1 0\or1\or2\or3\or4\or5\or6\or7\or8\or9\or
 A\or B\or C\or D\or E\or F\fi }

\makeatletter
\ifx\CUP@mtlplain@loaded\undefined
\else
\fi
\makeatother
 \makeatletter
 \ifx\CUP@mtlplain@loaded\undefined
   \font\tenbmi=cmmib10 at 10pt
   \font\sevenbmi=cmmib10 at 7pt
   \font\fivebmi=cmmib10 at 5pt

   \newfam\bmifam
   \textfont\bmifam=\tenbmi
   \scriptfont\bmifam=\sevenbmi
   \scriptscriptfont\bmifam=\fivebmi
   
 \fi
 \makeatother
\ifnfsstwo

\fi
\ifnfssone

\fi
\ifoldfss

\fi

\mathchardef\varLambda="0103

\makeatletter
\ifx\CUP@mtlplain@loaded\undefined
\else
\fi
\makeatother
\makeatletter
\ifx\CUP@mtlplain@loaded\undefined
  \font\tenbms=cmbsy10
  \font\sevenbms=cmbsy10 at 7pt
  \font\fivebms=cmbsy10 at 5pt
  \newfam\bmsfam
  \textfont\bmsfam=\tenbms
  \scriptfont\bmsfam=\sevenbms
  \scriptscriptfont\bmsfam=\fivebms

  \edef\bsy@{\hexnumber\bmsfam}
  \mathchardef\bnabla="0\bsy@72
\fi
\makeatother
\def\etal{\mbox{\it et al.}}

\title[The complete S5 polar cap sample]%
{The complete S5 polar cap sample:
en route to phase-delay global astrometry}

\author[E. Ros {\it et al.\/}]%
{E.\ns R\ls O\ls S$^1$,
J.\ls M.\ns M\ls A\ls R\ls C\ls A\ls I\ls D\ls E$^2$,
J.\ls C.\ns G\ls U\ls I\ls R\ls A\ls D\ls O$^{2}$,
M.\ls A.\ns P\ls \'{E}\ls R\ls E\ls Z-\ls T\ls O\ls R\ls R\ls E\ls S$^{3}$}

\affiliation{$^1$Max-Planck-Institut f\"ur Radioastronomie, Auf dem H\"ugel
69, D-53121 Bonn, Germany\\[\affilskip]
$^2$Departament d'Astronomia i Astrof\'{\i}sica, Universitat de
Val\`encia, E-46100 Burjassot, Spain\\[\affilskip]
$^3$ Istituto di Radioastronomia, CNR, Via P.\ Gobetti 109, I-40129 Bologna,
Italy}

\begin{document}
\ifnfssone
\else
  \ifnfsstwo
  \else
    \ifoldfss
      \let\mathcal\cal
      \let\mathrm\rm
      \let\mathsf\sf
    \fi
  \fi
\fi

\maketitle

\begin{abstract}
We report on the present status of our S5 polar cap phase-connected
astrometry program. 
We observe 13 radio sources in
the northernmost 20$^\circ$ of the sky
at the wavelengths of 3.6\,cm and 2\,cm, and we
plan to extend the program to 0.7\,cm.  We phase-connect jointly
all our data successfully.  We image the radio sources
and some of them show morphological changes, in which astrometric
registration is needed to determine the kinematics of the
source components. 
We aim at unprecedented astrometric accuracy and at a check of
the jet standard model at the 5-10\,$\mu$as/yr level.
\end{abstract}


IERS and USNO maintain the International Celestial Reference Frame (ICRF)
using at radio frequencies the VLBI group-delay observable.  
We aim to implement in 
the reference frame determination process the most precise VLBI observable: 
the phase-delay, which should improve astrometric
precisions at centimeter wavelengths.
A phase-connection is necessary to overcome the $2\pi$ ambiguous
nature of the phase-delay.
The final source position determinations are based on 
a weighted least-squares analysis 
of the connected phase-delays.

Recently, 
the phase-connection process has been extended to
sources 15$^\circ$ apart, with astrometric
precisions well below one milliarcsecond
(\cite{PerezTorres00}).
The avenue of mm-VLBI astrometry (with precisions higher than cm-VLBI)
is now also open, after 
\cite{Guirado00} successfully demonstrated phase-delay 
astrometry at {$\lambda$0.7\,cm} over
$5^\circ$.
Furthermore, 
the ionosphere can be modeled and removed from 
the astrometric observables from the results provided by the Global 
Positioning System
network (\cite{Ros00} and \cite{PerezTorres00}).

These developments encourage us to aspire to a more ambitious
program: phase-delay astrometry between 13 radio sources
to get accuracies
better than 0.1 milliarcseconds.  
We have an ongoing VLBA astrometric program
to test the absolute kinematics of components 
in the complete S5 polar cap sample.  It consists of the
BL-Lac objects 
0454+844, 0716+714, 1749+701, 1803+784, and 2007+777,
and the QSOs
0016+731, 0153+744, 0212+735, 0615+820, 0836+710, 1039+811, 1150+812,
and 1928+738.
It is possible to phase-connect all the data jointly by using
bootstrapping techniques:
these results are reported in \cite{Ros99b}.
To date we have carried out observations at four 
epochs, two at {$\lambda$3.6\,cm} 
({1997.93} and {1999.43}), 
and two at {$\lambda$2.0\,cm} ({1999.57} and 
{2000.46}).

\begin{figure}[tbh]
\vspace{264pt}
\includegraphics{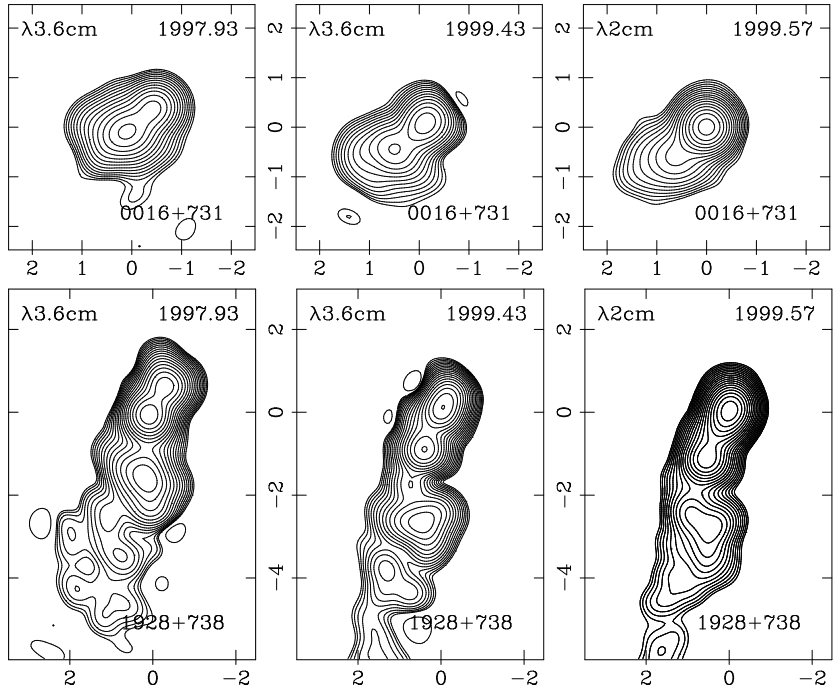}
\caption{Images of QSO\,0016+731 and QSO\,1928+738 selected from
our observations, convolved with a circular beam of 0.6\,mas diameter.
Contours are 2\,mJy$\times$$(-2,2,2\sqrt{2},4,\cdots)$.  Peaks of
brightness are --from left to right-- of 0.203, 0.162, 
and 0.442\,Jy/beam for 0016+731 and 0.961, 1.307, and 1.356\,Jy/beam
for 1928+738.
\label{fig:rosfig1}}
\end{figure}

In Fig.\ \ref{fig:rosfig1} we show 
VLBA images at different wavelengths for
two radio sources selected from our
observations.  For a better comparison, they are convolved with
the same beam and show the same contour levels.
Comparing the left and central panels for both sources, the peaks of brightness
belong to different features evolving in the jets.  
It is obvious that the image registration based on the position
of the peak of brightness (at the same observing frequency) 
is not correct, and that
an astrometric registration is needed
to identify components from one
epoch to another (left and central panels), and to 
perform spectral studies of the whole sample for nearby epochs 
(central and right panels).
This absolute
registration, only available from phase-delay astrometry, is 
needed to study the absolute kinematics of the components.

We plan to extend the 
astrometric monitoring of the 13 radio sources in the polar 
cap sample to 
{$\lambda$0.7\,cm}.
After several years, this program should provide higher
precision than any foreseeable astrometric program, and will
check the standard jet model in the complete sample at the 
{5-10\,$\mu$as/yr}
level.  Accurate registration of maps at the highest available resolution
will allow a study of jet components with unprecedented precision and
spectral information.  This project may be considered a first step
to extend phase-delay astrometry to the entire sky, and thus as a
complement, with a higher accuracy, to the
effort of ICRF.



%


\end{document}